\documentclass[10pt,conference,letterpaper]{IEEEtran}
\usepackage{times,amsmath,epsfig}
\usepackage{graphicx}
\usepackage{balance}  
\usepackage{algorithm}
\usepackage{amssymb}
\usepackage{algorithmic}
\usepackage{MnSymbol}
\usepackage{graphicx}
\usepackage{caption}
\usepackage{subcaption}
\usepackage{slashbox}
\usepackage{multirow}
\usepackage{color}
\usepackage{booktabs}
\usepackage{threeparttable}
\usepackage{wrapfig}
\usepackage{footnote}

\newcommand{\From}{\textbf{FROM}}
\newcommand{\Sketch}{\textbf{SELECT}}
\newcommand{\Group}{\textbf{GROUP BY}}
\newcommand{\Summarize}{\textbf{SUMMARIZE BY}}
\newtheorem{definition}{Definition}

\title{VCExplorer: A Interactive Graph Exploration Framework Based on Hub Vertices with Graph Consolidation}
\author{
{Huiju Wang $^{1}$ \hspace{0.6cm} Zhengkui Wang $^{3}$ \hspace{0.6cm} Kian-Lee Tan$^{2}$ \hspace{0.6cm} Chee-Yong Chan $^{2}$  \hspace{0.6cm} Qi Fan $^{2}$, \hspace{0.6cm} Xiao Yue $^{1}$}
\vspace{1.6mm}\\
\fontsize{10}{10}\selectfont\itshape
$^1$ School of Information  and Safety Engineering, Zhongnan University of Economics and Law, China 430073 \\
$^2$ School of Computing, National University of Singapore ,Singapore 117417\\
$^3$ InfoComm Technology, Singapore Institute of Technology ,Singapore 138683
}

\begin{document}
\maketitle
\begin{abstract} 
Graphs have been widely used to model different information networks, such as the Web, biological networks and social networks 
(e.g. Twitter). Due to the size and complexity of these graphs, how to explore and utilize these graphs has become a very 
challenging problem. In this paper, we propose, \bf{ VCExplorer}, a new interactive graph exploration framework that integrates the strengths of graph visualization and graph summarization. Unlike existing graph visualization tools where
vertices of a graph may be clustered into a smaller collection of super/virtual vertices, VCExplorer displays a 
small number of actual source graph vertices (called {\em hubs}) and summaries
of the information between these vertices. We refer to such a graph
as a HA-graph (Hub-based Aggregation Graph).
This allows users to appreciate the relationship between the hubs, 
rather than super/virtual vertices. 
Users can navigate through the HA-graph by ``drilling down'' into the summaries between hubs to display more hubs. 
We illustrate how the graph aggregation techniques can be integrated into the exploring framework as the consolidated information to users. In addition, we propose efficient graph aggregation algorithms over multiple subgraphs via computation sharing. Extensive experimental evaluations have been conducted using both real and synthetic datasets and the results indicate the effectiveness and efficiency of VCExplorer for exploration.

%
\end{abstract}

%
\section{Introduction}
Graphs are powerful tools to model a variety of information networks, such as the Web, biological networks and social networks (e.g. Twitter). In a graph, each vertex usually represents one real world object and each edge indicates the link between two objects. Normally, both vertices and edges may be annotated with attributes or labels. 

These graphs contain a wealth of valuable information to support a wide variety of queries for information discovery and decision making. To better understand the information encoded in the underlying graphs, different approaches have been used to explore these data. 

On one hand, we have {\em summarized-based} methods that aim to simplify 
or summarize the 
graphs to provide a coarser and higher level graph that is normally 
referred to as a view. These approaches include graph summarization~\cite{Tian2008}, graph aggregation in graph OLAP~\cite{Zhao2011}, graph clustering and so on. 
The common methodology of these approaches is to aggregate multiple vertices (resp. edges) into one super node (resp. edge) based on certain 
rules (e.g. through clustering or aggregating the vertices with the same attributes) 
to a view with much fewer vertices and edges. This makes it easier to 
visualize a large and complex graph. On the other hand, 
we have {\em graph-based} methods (e.g. \cite{surveyvisualization})
that convey the content of a graph 
by displaying the 
whole graph including all the individual vertices and the links on 
a screen via graph layout. The mainstream approach of these mechanisms
is graph visualization which provides the individual vertices and the 
links among them in the visualization space.

From users' point of view, graph summarization/aggregation methods
show summarized view, but hide the original individual vertices; 
conversely, graph visualization schemes show all individual vertices, 
but hide the summarized view. Each of the approaches has its own strengths 
and limitations in exploring a graph. As the size of the graph increases, 
what to show and what to hide plays an important role 
on the effectiveness of graph exploration.

\subsection{A Running Example over Social Network.} 

\begin{figure*}
\vspace{-12pt}
\centering
\includegraphics[width=5.6in]{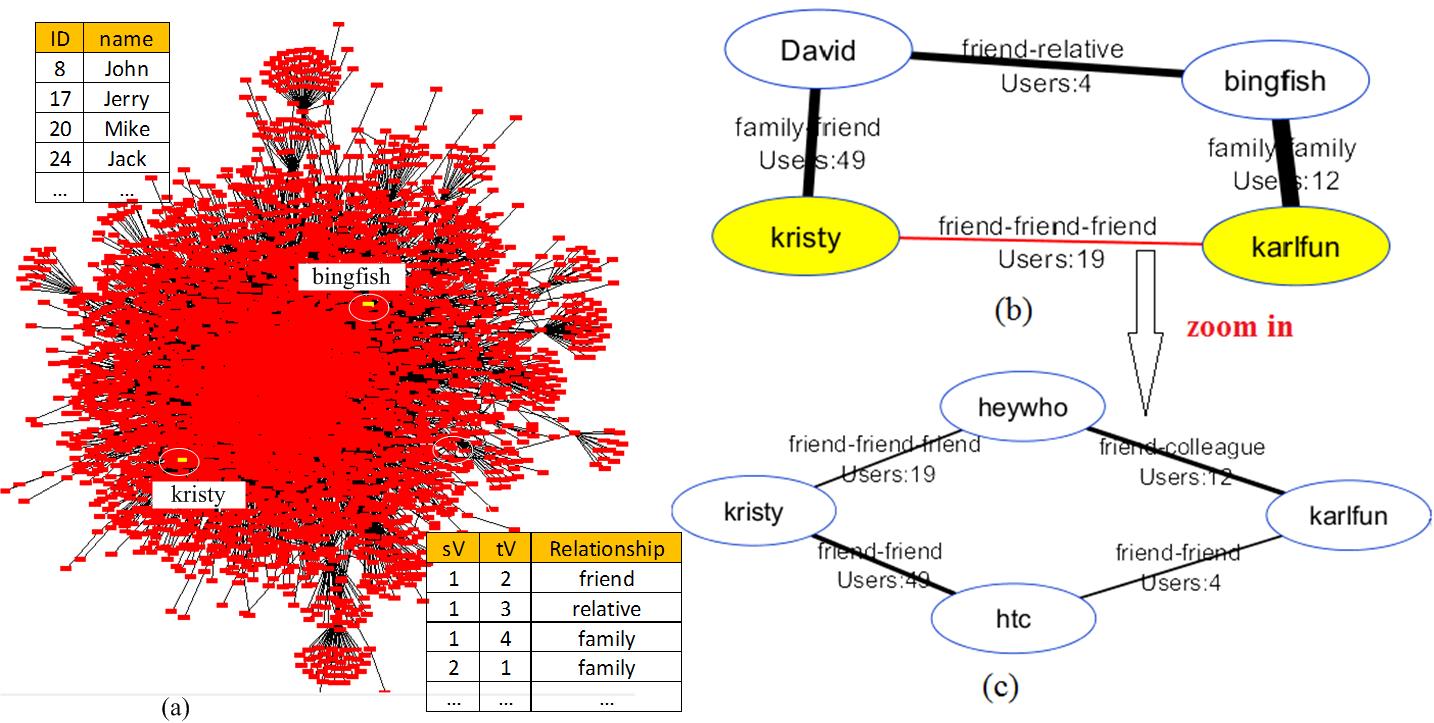}
\caption[Social Network Example]{A running example of VCExplorer. (a) A derived Twitter network dataset \footnotemark{}
with 5k vertices and 18k edges visualized by Cytoscape~\cite{UIUCDataset} 
(b) output HA-graph of SQ1. 
(c) HA-graph after zooming in edge ($kristy$, $karlfun$) in (b). 
In (b) and (c), the  width of an edge represents the relationship strength of the induced subgraph represented by the edge; 
and each edge is labeled with its representative relationship type as well as a count of the number of vertices in the associated induced subgraph.} 
\label{fig:socialnetwork}
\end{figure*}

Typically, a social network is modeled as a graph. Vertices of the 
graph represent persons, whereas edges represent relationships between 
the vertices. Both vertices and edges may have attributes. 
Figure~\ref{fig:socialnetwork}(a) shows such a social network. 
Each vertex is affiliated with an attribute {\em name}, 
and each edge is affiliated with a {\em relationship type} (e.g., friend, relative)
between two vertices. Given such a social network, an analyst may be interested to find out \textit{how user bingfish is connected with user kristy}. 
Now, each path between {\em bingfish} and {\em kristy}
represents one type of connections between them,
and there are potentially an exponential number of such paths. 
Under the graph-based methods, it is not feasible to show the entire
graph (or the subgraph containing all paths between them)
to users as the display will become too cluttered
(as shown in Figure~\ref{fig:socialnetwork}(a)).
With summarized-based methods, the resultant view resulted in
information ``loss'' - the vertices of $bingfish$ and $kristy$ are not
shown at certain levels. Therefore, for the aforementioned query, 
both approaches cannot effectively facilitate exploration.

In this work, we advocate an alterative approach that   
displays a subgraph (called HA-graph) containing
a subset of the actual vertices (called hubs)
between {\em bingfish} and
{\em kristy}\footnote{Note that both vertices {\em bingfish} and
{\em kristy} are also hubs.} 
as well as summaries of the relationships and
information between these vertices. 

\footnotetext{The network is consisted of  bi-directional edges of the input Twitter network. For clearness, we draw bi-directional edges as undirectional ones in Figure~\ref{fig:socialnetwork}}
Such an approach allows users to be engaged with the original/source vertices 
(rather than virtual vertices), and the consolidated summary
information of the {\em hidden} vertices (i.e, vertices that are not
hubs in the current graph).  Our approach may be viewed as a generalization
of the above two approaches: if all vertices are chosen as hubs, 
it becomes a graph visualization approach; if no hub is selected, 
it becomes a graph summarization approach. 
We have developed VCExplorer (\textbf{V}ertex and \textbf{C}onsolidation Based \textbf{Explorer}),
a novel graph exploration framework that does just precisely what we
advocate.
VCExplorer starts by accepting a new type of graph exploration
query (denoted as GE-query) that is formally 
defined in Section~\ref{sec:framework}. 
The following is an example GE-query, denoted by \textbf{SQ1}, on the social network graph $G$ in Figure~\ref{fig:socialnetwork}:

\begin{flushleft}\tt
\hskip 10pt \Sketch\ TopMaxDegreeVertices(G', 2)\\
\hskip 10pt \From\ Subgraph(G, kristy, bingfish, 4) G'\\
\hskip 10pt \textbf{GROUP BY} betweenness() \\
\hskip 10pt \Summarize\ relationshipStrength(),  \\
\hskip 88pt relationshipType(),\\
\hskip 88pt vertexCount()
\end{flushleft}

Given a GE-query, VCExplorer first derives the target subgraph to 
be explored. For social network applications, we expect users to 
explore relationships among people close to each other. 
In \textbf{SQ1}, the \From\ clause specifies the subgraph of interest to be explored
by using a user-defined function {\em Subgraph},
which extracts the subgraph $G'$ of $G$ that consists of all vertices/edges along paths (with a path length of at most 4 hops apart) between a specific pair of vertices,
$bingfish$ and $kristy$\footnote{If $bingfish$ and $kristy$ are 
more than 4 hops apart, then we should use that distance to bound the search space.}. 
The \Sketch\ clause identifies a set of hubs using a user-defined function,
{\em TopMaxDegreeVertices(G', 2)},
which returns a set of two vertices in $G'$ with the maximum vertex degree;
these  hubs represent the two most influential people connecting $bingfish$ and $kristy$.
For \textbf{SQ1}, suppose that $David$ and $karlfun$ are the top 2 vertices selected.
Unlike graph visualization methods, only the hubs will be displayed in the resultant graph (as shown in Figure~\ref{fig:socialnetwork}(b)).
In this way, it is visually more appealing since fewer but more important vertices are being displayed. 

Given the hubs (including vertices {\em kristy} and {\em bingfish}), 
the \textbf{GROUP BY} clause then induces a subgraph of $G'$ between every pair of the hubs
using a user-defined function which determines
for each induced subgraph $G'(x,y)$ (wrt a pair of hubs $x$ and $y$ in $G'$) and for each vertex $v$ in $G'$,
whether $v$ is contained in $G'(x,y)$.
For \textbf{SQ1}, the {\em betweenness} function
in the \textbf{GROUP BY} clause 
includes a vertex $v$ in an induced subgraph $G'(x,y)$
if $v$ is along some path between $x$ and $y$ in $G'$. One edge belongs to G'($x$, $y$) if its two vertices are in G'($x$, $y$). 
Note that a vertex/edge could be contained in multiple induced subgraphs.

The \Summarize\ clause specifies a list of user-defined aggregation functions to compute summary information for each of the induced subgraphs.
In \textbf{SQ1}, the user is interested in the following three summary information for each induced subgraph $G'(x,y)$.
The first is the closeness of the two hubs $x$ and $y$ based on the trust propagation among the users in $G'(x,y)$~\cite{XLin2014} computed by the {\em relationshipStrength} function.
The second is the most representative relationship between the two hubs, such as ``friend's friend'' relationship; 
the {\em relationshipType} function returns the concatenation of the relationship types along the shortest path between $x$ and $y$.
The third is a count of the number of vertices in the induced subgraph $G'(x,y)$ which is computed by the {\em vertexCount} function.

In general, all the information discovered can be visualized as a graph, referred as a \emph{Hub-based Aggregation Graph} (HA-graph) in this paper. 
In the resultant HA-graph, the vertices are the hubs and 
edges are the connections among them which will be associated with the 
summarized information. For instance, the resultant HA-graph of \textbf{SQ1} is shown in Figure~\ref{fig:socialnetwork}(b). The HA-graph is 
much clearer than visualizing all the vertices in the underlying graphs. 
In addition, the HA-graph allows users to 
navigate and explore
by zooming to the next level. To analyze the reason why $kristy$ and 
$karlfun$ is weakly connected, the analyst may zoom in to the subgraph between 
$kristy$ and $karlfun$ by issuing another GE-query. 
The resultant graph is shown in Figure~\ref{fig:socialnetwork}(c).

\subsection{Contributions}

Our contributions may be summarized as follows:
\begin{itemize}
\item{
We present VCExplorer, a novel graph exploration framework.
VCExplorer combines the innovative ideas of graph visualization and graph 
summarization. On one hand, it shows a subset of vertices each time without
cluttering the display; and on the other hand, it summarizes information
of ``hidden'' vertices.  
Compared to traditional graph visualization approach, VCExplorer 
is able to provide much clearer and useful information. It also offers an effective mechanism to navigate through the graph.
}

\item {We illustrate how VCExplorer framework can be designed by incorporating existing technologies. Each component of VCExplorer actually covers many research problems and most of them have been studied for a long time. We further study how the newly emerged graph aggregation can be well integrated with the VCExplorer as one approach to summarize the relationship between two hub vertices. We propose and study efficient algorithms to share computations to salvage partial work done. }

\item{We conduct extensive experimental evaluation based on both real and synthetic data. The experimental results demonstrate that VCExplorer is effective and efficient.}

\end{itemize}

\section{VCExplorer: The Big Picture}
\label{sec:framework}
It is interesting and challenging to develop techniques to support graph
exploration in real-time. In this section, we introduce VCExplorer by
giving an overview of its features and components. 

\subsection{Graph Exploration Query}
\label{sec:squery}

The exploration starts by accepting a user's query defined as follows.

\begin{definition}
A graph exploration query (GE-Query) is used to explore a data graph $G$ 
by identifying a subgraph $G'$ of interest, a subset of interested vertices (i.e. hubs) in $G$',
and computing summarized information for each subgraph induced by every pair of hubs in $G'$.
A GE-Query is characterized by five components  $(G, \pi, \sigma, \gamma, \{\tau_1,\cdots,\tau_n\})$
which can be expressed using the following syntax:
\begin{flushleft}\tt
\hskip 20pt \Sketch\ $\sigma(G')$ \\
\hskip 20pt \From\ $\pi(G)$ $G'$\\
\hskip 20pt \Group\ $\gamma(G',x,y)$\\
\hskip 20pt \Summarize\ $\tau_1(G'(x,y))$,$\cdots$,$\tau_n(G'(x,y))$
\end{flushleft}
\label{def:sketchquery}
\end{definition}
where
\begin{itemize}
\item $G$ is an input data graph from which a subgraph $G'$ of interest is extracted from a user-defined function $\pi()$.
\item $\sigma$ is a user-defined function to return a set of hubs from the subgraph of interest $G'$.
Possible selection criterias for $\sigma$ include ``\emph{selecting vertices with a specific attribute value}", 
``\emph{selecting the top k vertices with maximum closeness centrality value}" and so on. 
For each selection criterion, the system may build an index  to accelerate the computation of the selection. 
\item $\gamma$ is a user-defined function to compute an induced subgraph of $G'$,
denoted by $G'(x,y)$,
for each pair of hubs $(x,y)$ from $\sigma(G')$.
An example of $\gamma$ is the {\em InBetween} function illustrated in \textbf{SQ1}, whose computation can be accelerated using some reachability index. 

\item Each $\tau_i$ is an aggregation function to compute some summarized information for each of the induced subgraphs $G'(x,y)$.
The summarized information could be path-related information (e.g., shortest path length), 
aggregation information (e.g., aggregate graph based on different attributes like in graph OLAP \cite{Zhao2011}). 
In~\cite{hjwang2014}, we have developed aggregation sharing algorithms by utilizing the overlaps between subgraphs to share computations.
\end{itemize}

\subsection{Hub-based Aggregation Graph}
\label{subsec:sketchgraph}

The output of a GE-query is formally defined as a HA-graph.

\begin{definition}
\label{def:sketch_graph}
Hub-based Aggregation Graph (HA-graph): Given a GE-query 
$(G, \pi, \sigma, \gamma, \{\tau_1,\cdots,\tau_n\})$,
the result is a graph called the HA-graph 
$H$ = ($V$, $E$),
where $V$ is the set of hubs extracted from the subgraph of interest $\pi(G)$;
note that the set of hubs also include any vertex argument in $\pi$ function for computing the subgraph of interest.
$E = \{(x,y) |\ x,y \in V, \gamma(\pi(G),x,y)$ is a non-empty graph$\}$.
Each vertex $v \in V$ is associated with a set of attribute values inherited from the corresponding vertices in $G$.
Each edge $(x,y)$ in $E$ is associated with a set of summarized values $\{t_1,\cdots,t_n\}$
where each $t_i = \tau_i(\gamma(\pi(G),x,y))$ is an aggregated value computed by the aggregation function $\tau_i$ on the induced subgraph for the pair of hubs $(x,y)$.
\end{definition}


Figure~\ref{fig:socialnetwork}(b) shows the resultant HA-graph for \textbf{SQ1}, 
which consists of two most influential users between $kristy$ and $bingfish$. 
The labeled edges between a pair of vertices indicate the summarized information for the induced subgraph betwen the vertices. 
For instance, the edge ($kristy$, $karlfun$)
in Figure~\ref{fig:socialnetwork}(b) indicates that the number of vertices in the induced subgraph between $kristy$ and $karlfun$ is 19, the shortest path between them in the induced subgraph is 3 consisting of three $friend$ edge labels along this shortest path, and they have weaker relationship strength comparing with other pairs of hubs.

\subsection{Navigation}

It is essential to provide navigation capabilities in graph exploration. 
This is to allow users to interact and explore large graphs. 
In general, zooming operations are quite indispensable and useful. 
Given a HA-graph, users can zoom-in on an induced subgraph $G'(x,y)$ by clicking on its corresponding edge $(x,y)$.
Another way for users  to zoom-in is to select a subset of the vertices in the HA-graph; 
the collection of induced subgraphs among the selected vertices would form a new subgraph of interest to be further explored.


\section{Framework Design}
After defining VCExplorer framework, we turn to the design of such framework. Specifically, we discuss how to utilize existing techniques to design efficient algorithm for each component. Due to space limit, we will not go too far into the technical details.

\subsection{Hub Vertex Generation}
\label{sec:svg}
Hub vertices are selected using the $\Gamma$ function which is based on some \emph{measures}, such as vertex attribute, importance values, etc. According to the variability of measure value, we classify them into two categories: 

\textbf{Static function}: whose measure values does not change during subgraph navigation. Such measures include vertex attributes and derived attributes. Take Twitter network as example.  $\Gamma$ function \emph{"Users whose age is above 80"} takes \emph{age} as measure which is an attribute native to vertex and remain static during navigation. $\Gamma$ function \emph{"Top-10 Americans rank with closeness centrality value in ascending order"} is built on \emph{closeness centrality}.  As \emph{closeness centrality} measure is defined in the context of whole graph, during navigation, the \emph{closeness centrality} value do not change in the context of new subgraphs. In this context,  closeness centrality is in fact a derived attribute for vertex.

Since this kind of measures are static, it can be precomputed (for derived attribute) and indexed. For example, we can pre-compute the \emph{Twitter Closeness Centrality} for every vertex and index them using $B^+$-tree. When processing $\Gamma$, we can directly refer to the $B^+$-tree for a candidate list and thus boost the $\Gamma$ computation. 

\textbf{Dynamic Function}: whose measure values change accordingly during subgraph navigation. For example, given a $\Gamma$ which computes \emph{"Top-10 Americans rank with Closeness Centrality in ascending order"}, here \emph{Closeness Centrality} measure implicitly refers to current subgraph that consists of American users and following relationship between these users. During navigation, since subgraph is changing, the \emph{Closeness Centrality} also changes.

Dynamic measures are often not easy to index, a commonly used technique is to compute the measure at run-time. When online computing is time consuming, we generally have two alternatives: 1) use approximated measure to speed up. For example in the case of \emph{Closeness Centrality}, we can adopt the approximate scheme as in \cite{BaderKMM07, Eppstein2001}. 2) precompute some intermediate results. In the case of \emph{Closeness Centrality}, we can compute all-pair shortest distance first. Since during navigation, subgraphs are extracted based on reachability property, all shortest distances are valid locally. With the knowledge of shortest distance, the \emph{Closeness Centrality} can be efficiently computed.

\subsection{Subgraph Extraction} 
Before consolidation, subgraphs between any pair of hub vertices are extracted. By default, a \emph{betweenness} function is used. That is for a vertex $v$ and two hub vertices $sv_1,sv_2$, $v$ is between $(sv_1,sv_2)$ iff $sv_1 \leadsto v$ and $v \leadsto sv_2$. Given an exploring graph $G$ and a set of hub vertices $SV$, subgraph extraction can be translated as:
$\forall v \in G$, compute two sets: $S(v) = \{sv| sv \in SV \wedge sv \leadsto v\}$, $R(v) = \{sv | sv \in SV \wedge v \leadsto sv\}$. The Cartesian product of $S(v)$ and $R(v)$ denotes the set of subgraphs $v$ belongs to.  Reachability index can be used to boost the extraction process. If the index is built and extraction is based on betweenness measure, we can use the following approach:

\textbf{Index based extraction}: Given a reachability index, the extraction can be performed as follows: $\forall v \in G$  $\forall sv \in SV$, conduct two reachability tests $v \leadsto sv ?$ and $sv \leadsto v?$. And then update its $S$ and $R$ lists accordingly. Many reachability indices are developed in literature, such as transitive closure, 2-hop \cite{Cohen2002}, highway \cite{Jin2012}, dual-labeling \cite{Wang2006} etc. Due to betweeness measure, it is easy to see that reachabiilty relationship holds in any subgraphs. Therefore, the index can be reused in further navigation.If the reachabality index is unavailable, we can adapt graph traversal based approach instead.

\textbf{Non-Index based extraction}: We first preprocess the graph to assign each vertex $v$ with a topological order number. Circles are condensed and vertices in the same circle share the same order number. Then we process the vertices topologically. For every vertex, it will push its $S$ lists to all its immediate children. Each children unions all the $S$ lists it received from its father to form its own $S$. The procedure to compute $R$ lists is similar but in a reversed manner. By so doing, the subgraph is extracted, but is slower than index based approach.

\subsection{Consolidation} 
After subgraph extraction,  consolidation is performed on each subgraph. According to object type to be consolidated, graph consolidation can be further classified into following categories:

\textbf{Attribute-based consolidation}: consolidation that is only operate on vertices (edges) attributes or derived attributes. Typical operators are SUM, COUNT, AVG, etc.  Since all the vertices (edges) are known at this stage, we can retrieve related attributes from the vertex (edge) attribute table. If any index on vertex(edge) ID is present, we can directly retrieve the target attributes, otherwise one scan on vertex (edge) attribute table will be introduced.

\textbf{Structure-based consolidation}: consolidation that is only related to graph structure. Typical operators include shortest distance (path), minimum cut etc. These problems are well-studied in the literature. Taking shortest distance as example, we have several algorithms to choose from: 1) In uniweighted graph, a BFS from a hub vertes is sufficient to compute all the shortest distance to other hub vertices; 2) In weighted graph, a Dijkstra's algorithm  is applicable; 3) If shortest distance indices \cite{wei2010tedi,xiao2009efficiently} is available and  subgraphs are extracted based on $betweenness$ function, the distance can be efficiently derived. 

\textbf{Attribute and Structure based consolidation}: consolidation that is related to both graph structure and attributes on vertex (edge). Prominent example in this category is graph aggregation. Several algorithms has been developed recently \cite{wangzk2014,Zhao2011}.Since these schemes focus on single graph computation, one naive solution is to run these schemes for each subgraph. Unlike the above two category where proper indices can boost the consolidation, graph aggregation is more complex and no indexing scheme is available. In the next Section, we will give an efficient algorithm to perform graph aggregation on multiple graphs.

\subsection{Visualization}
A HA-graph usually has at most tens of vertices, and hundreds of edges, thus most layout algorithms~\cite{herman2000graph} is able to handle it. In additional to displaying HA-graph structure, we also display the consolidated information for each edges in the HA-graph. A consolidated information can be a single value (i.e., COUNT(.)), a list (i.e., a shortest path, a set of group-value pairs) or a attributed graph (i.e., an aggregate graph). Given the diversity of the information, we create two modes for displaying the information on edges. 

Data Mode: we display results in raw data format. A single value is a label to an edge; A path is displayed as a list of vertices and is attached to edge. A graph is displayed as 2-D tables with each row representing an edge or a vertex in aggregated graph.

Graph Mode: we display results in graph format. A single value is still a label; A path is displayed as a chain; and A graph is displayed in vertex-edge format.

Users are flexible to toggle between two modes for each edge. We adapted several other interaction designs which are more friendly to users. In HA-graph, user can enable hover features, then all results on edges are hidden and only shown when mouse is moving over. In data mode, user are freely to perform selection, projection, sorting on 2-D table.

\section{Sharing-based Online Aggregation}

Efficient online aggregation for multiple subgraphs is the key to provide user better exploration experience. In this section, we introduce how to conduct the graph aggregation for multiple subgraphs online. We first introduce some preliminaries followed by one naive aggregation algorithm - SN (Shared Nothing) algorithm. Then we introduce the proposed aggregation algorithm, AS (Aggregation Sharing) algorithm, that provides an efficient aggregation by sharing the computation. 


Graph aggregation offers a high level view of the attribute graph~\cite{wangzk2014}. Integrating graph aggregation with VCExplorer is of great help to provide users the summarized information of the subgraphs which are unable to display. In this work, we will focus on the discussion of the distributive and algebraic functions (e.g. SUM, COUNT, Max, Min etc.) which can be applied to the subset of the edges or vertices in one graph. For these functions, the final results can be further calculated based on the result of each subset. For illustration, we take directed graph in Figure~\ref{fig:motivationexample} as input graph and use betweenness function to find out target subgraphs one vertex belongs to.  Other types of aggregation functions and graphs should be addressed similarly.

\subsection{Preprocessing}
\subsubsection{Handling SCC}

For directed graph, when betweenness is set chosen as the manner to extract the influential subgraph between two hub vertices, once one of the vertices in a SCC (strongly connected component) is in the subgraph, the entire SCC will be in the subgraph. Therefore, one optimization can be adopted here is to preprocess each SCC in advance.

In graph aggregation, each SCC can be pre-aggregated together and condensed into a super vertex. The super vertex will be associated with the pre-aggregate value of the SCC. In so doing, the original graph becomes an acyclic graph which is our discussion focused on in the later. Note that many existing works have been proposed and can be adopted here to detect the SCC, such as the Tarjan's strongly connected component algorithm that runs in $O(V+E)$.

\subsubsection{Tags Generation}
    For illustration, we first definition vertex and edge tags which will be used later. In $G$, every vertex is associated with a conceptual tag indicating which influential subgraph it belongs to in the HA-graph $G_s$.
\begin{definition}
Vertex Tag: $T(v)$ is a tag for every $v \in$ G. $T(v) = S(v) \bullet R(v)$, where $S(v) = \{ u | u \in G_s \wedge u \leadsto v \}$ and $R(v) = \{u | u \in G_s \wedge v \leadsto u\}$.
\end{definition}

\begin{figure}[th]
\centering
\includegraphics[width=3.5in]{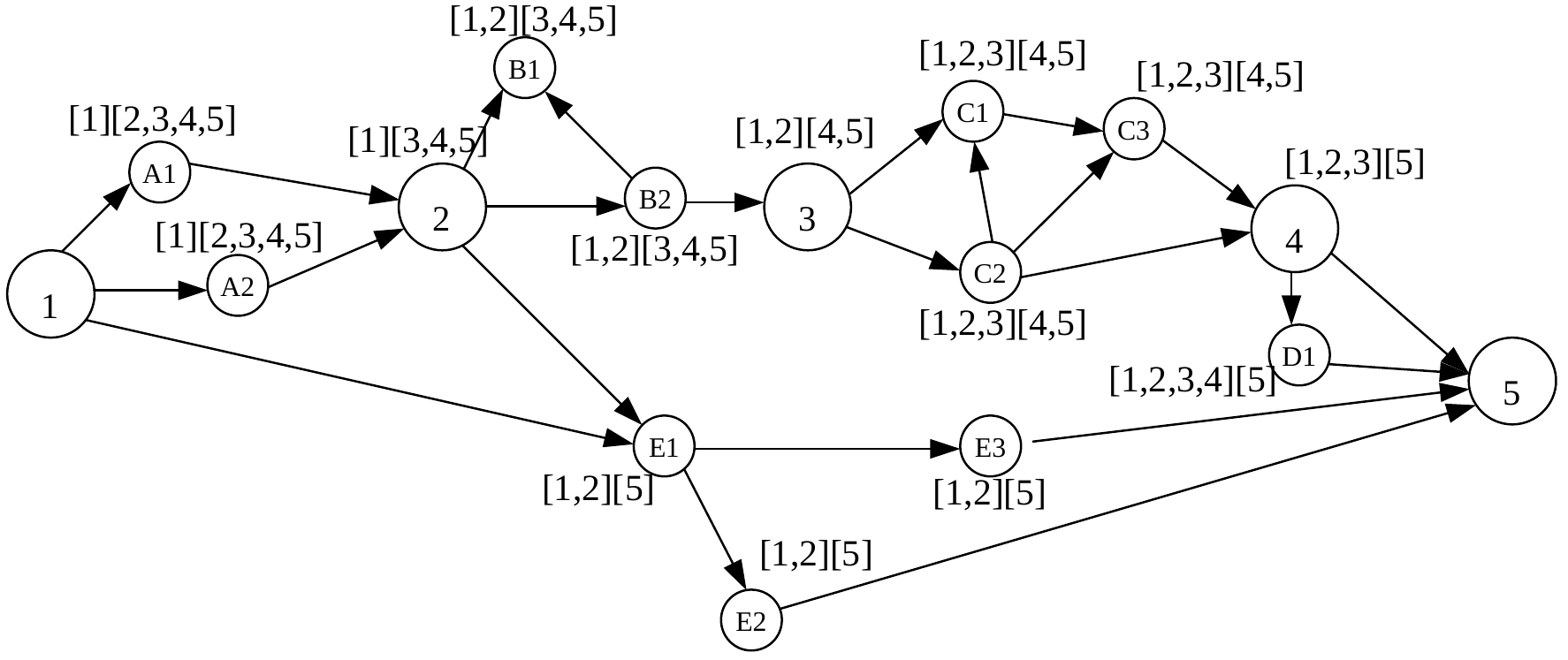}
\caption{Example Graph.}
\label{fig:motivationexample}
\end{figure}

Intuitively, $S(v)$ denotes the hub vertices which can reach $v$ in $G$ and $R(v)$ denotes the hub vertices which can be reached by $v$ in $G$. $T(v)$ is formed by concatenating the two lists. For instance, Figure \ref{fig:motivationexample} indicates a simple graph where vertices 1, 2, 3, 4 and 5 are selected as the hub vertices. In this example, vertex $A_1$'s tag is $<1><2,3,4,5>$, which means vertex 1 can reach $A_1 $ and $A_1$ can reach $<2,3,4,5>$. We refer to the list of S(v) as $T_S(v)$ and $R(v)$ as $T_R(v)$.

\begin{figure}[thp]
\centering
\includegraphics[width=3.5in]{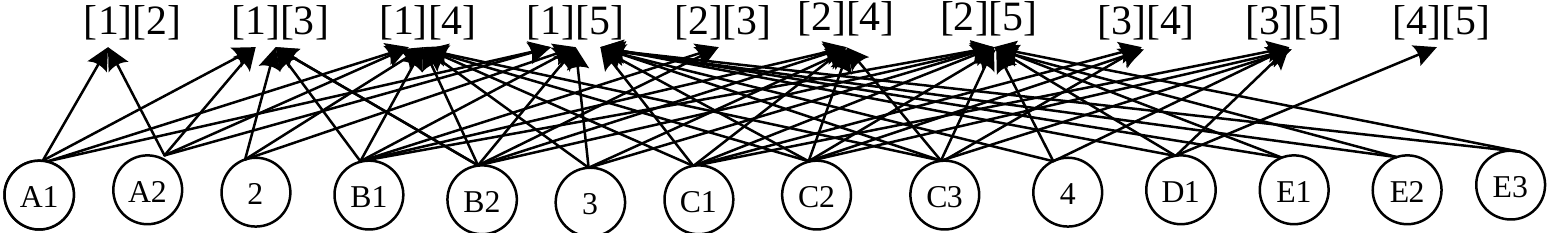}
\caption{SN-Agg Plan Example}
\label{fig:sn-agg}
\end{figure}

On the basis of tag definition, given a tag $T(v)$ of $v$, Cartesian product of $T_S(v)$ and $T_R(v)$ represents the infulential subgraphs $v$ belongs to. In addition, we also define the the size of the Cartesian product as the cardinality of $T(v)$.
For instance, in Figure \ref{fig:motivationexample}, $A_1$ is tagged with $<1><2,3,4,5>$ indicating that it belongs to subgraphs $<1,2>$, $<1,3>$, $<1,4>$ and $<1,5>$ and the cardinality $|T(A_1)|$ is 4.

Similarly, we assign the similar tag for edge tag as well. In $G$, a tag for $e(s,t)$ is denoted as $<T_S(s), T_R(t)>$. For instance, $e(c1, c3)$ is tagged with $<T_X(c1), T_R(c3)>$ (<1,2,3><4,5>). The cardinality of $e(c1, c3)$ is 6.


To speedup generating the tags, the reachability index can be adopted here, such as transitive closure or 2-hop. For each $v \in G$ and $u \in G_S$, we test whether $v \leadsto u$ or $u \leadsto v$. The total complexity is $O(|V|*k*r)$, where $k$ stands for the number of hub vertices and $r$ stands for the cost for reachability testing between two vertices. After generating the tags for each vertex, the edge tags can be easily calculated based on the vertex tags.

\subsection{Share-Nothing Aggregate Algorithm}
\label{subsec:SN}

Recall that there are multiple subgraphs need to be aggregated, each of which corresponds to one edge in $G_s$. To conduct the graph aggregation, one naive approach is to aggregate each subgraph individually. Intuitively, this approach aggregate the subgraph independently without any sharing operation. Thus, we refer to this algorithm as $SN$ algorithm - stands for shared nothing.

In \emph{SN} algorithm, each subgraph extracts its own vertices and edges and further calculates its own aggregate graph independently. Take the vertex aggregation as in example. Figure \ref{fig:sn-agg} shows how the vertices will be processed for different subgraphs. In Figure \ref{fig:sn-agg}, the bottom lists all the vertices and the top lists all the subgraphs. Each link between the vertex and subgraphs indicates one aggregate operation where the vertex should be aggregated to a corresponding subgraph. Thus, in $SN$, each subgraph (denoted by tags) receives and aggregates the vertices independently. Given a graph with $n$ vertices, assume that $|S|$ is the number of subgraphs, the complexity of vertex aggregation is $O(m|S|)$.

For the edge aggregation, if the graph is stored in the format as shown in Figure \ref{fig:socialnetwork}, there is a need to convert the vertex IDs of two endpoints of one edge to the vertex aggregate attributes. This can be done by performing a join between the edge attribute table and vertex attribute table. After the conversation, the edge aggregation can be conducted in the similar way as the vertex aggregation. Given the a graph with $m$ edges and $|S|$ subgraphs, the complexity of edge aggregation is $O(m|S|)$.

\subsection{Aggregation Sharing Algorithm}
\label{subsec:AS}
\emph{SN} is a straightforward approach as it computes the graph aggregation for each subgraph independently. However, it may incur high computation overhead as it may involve many redundant computations. 

One observation is that some vertices and edges are involved the same set of multiple subgraphs. This provides us the opportunity to share the computation among different subgraphs.

\begin{figure}[thp]
\centering
\includegraphics[width=3.2in]{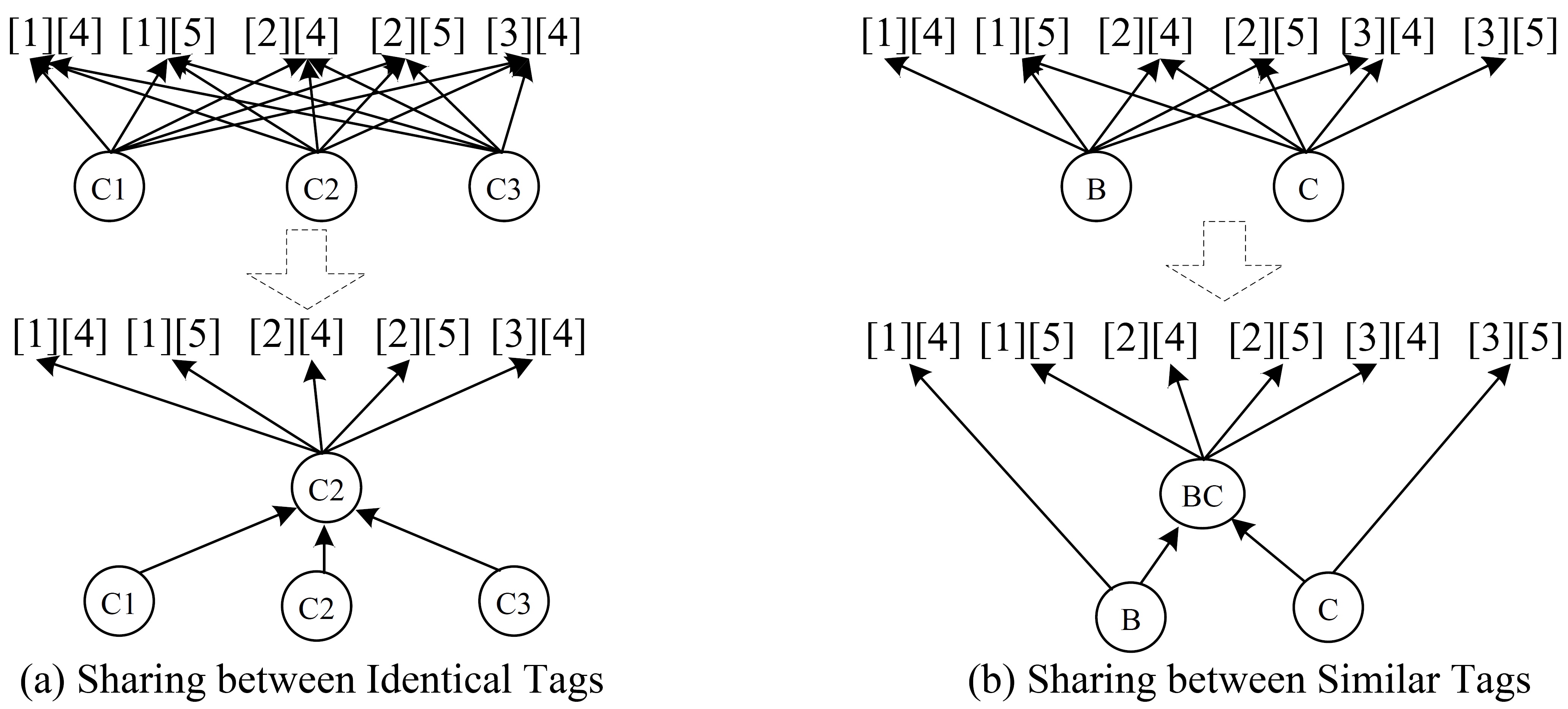}
\caption{Sharing Plan Example}
\label{fig:share-example}
\end{figure}

For instance, in Figure~\ref{fig:sn-agg} $C_1, C_2, C_2$ have common tag of $<1,2,3><4,5>$ which means these three vertices are involved into the same 6 subgraphs $<1,4>, <1,5>, <2,4>, <2,5>, <3,4>$ and $<3,5>$. Therefore, the aggregation computation can be shared among these subgraphs. $C_1, C_2, C_2$ can be aggregated once and then supply to the 6 subgraph directly, instead of aggregating them 6 times. Similarly, $B_1, B_2$ can also be aggregated together then supply the result to their shared subgraphs directly. Figure~\ref{fig:share-example} (a) indicates this procedure where B (resp. C) is the aggregate result of $B_1$ and $B_2$ (resp. $C_1, C_2, C_2$). 

Another observation is that even though the tag are not exactly the same, they may still be able to share the computation once they have the shared subgraphs. One simple example is between B ($<1,2,3><4,5>$) and C ($<2,3><4,6>$) which are similar but not the same. It is easy to see that they share 3 subgraphs $<2,3><4>$. We can pre-aggregate B and C where the result can be directly supplied to the 3 shared subgraphs which is able to reduce the computation overhead. Figure~\ref{fig:share-example} (b) indicates such an idea.

Based on these observations, we propose a new algorithm, \textbf{AS (Aggregation Sharing)}, on the principle of sharing the aggregation when the vertices or edges are involved into a common set of subgraphs. We refer to a common set of subgraphs as a shared component(SC). Given two tags t1 and t2, the SC can be calculated by t1.S $\wedge$ t2.S concatenated by t1.R $\wedge$ t2.R where $\wedge$ means intersect. For instance, given t1 ($<1,2,3><4,5>$) and t2($<2,3><4,6>$), the SC can be calculated as $<1,2,3> \wedge <2,3>$ concatenated by $<4,5,> \wedge <4,6>$ which will be $<2,3><4>$. Note that to speed up SC calculation, the vertex ID lists in the tag can be stored as BitSet where the SC can be simply computed via the AND operation between two BitSets.

\textbf{AS Algorithm:} Discovering all the possible SCs among the tags incurs a high computation complexity that is almost $2^n$ where n is the number of different tags. As a real-time exploration, finding the optimal solution for finding SCs may not be practical. Therefore, in this paper, we propose a heuristic algorithm to discover the SCs by tag clustering. For illustration, as the aggregating the vertices and edges is under the similar procedure, we focus on introducing the vertex aggregation here. The similar algorithm can be easily adopted for the edge aggregation which will be omitted. The pesudo code of proposed \textbf{AS} algorithm is provided in Algorithm \ref{alg:aggregateplan}. 

\begin{algorithm}[th]
\caption{Aggregation Sharing Alogrithm}
\begin{algorithmic}[1]
\STATE \textbf{INPUT:} $vertices$
\STATE $aggPlan$:=null;
\STATE $clusters$:=null;
\STATE $tags$:= genTags(vertices)
\label{code:gentags}
\STATE $queue$:= sort($tags$)
\label{code:tagsort}
\WHILE{!$queue.isEmpty$}
\label{code:secondbegin}
		\STATE $nt$:= $queue$.pop()
		\IF{nt is the same tag with previous one}
		\STATE Combine nt into current group $g$
		\label{code:combine}
		\ELSE
		\IF{$|tag| \ge 3 $}
			\STATE $cluster$:=FindBestCluster($g$, $clusters$)
			\label{code:findCluster}		
			\STATE $ct$:=$cluster.tag$
			\IF{Saving of combing $g$ and cluster is positive}		
				\label{code:addtag}	
				\STATE $cluster$.add($g$)
				\STATE $st$:= $ct \wedge nt$
				\label{code:save1}
				\STATE $nt_s$:=$nt - s$
				\STATE $ct_s$ :=$ct -s$
				\STATE $ct$:=$st$
				\label{code:save2}
				\STATE $aggPlan$.add($ct_s,cluster$)
				\label{code:addcts}
				\STATE $aggPlan$.add($nt_s, aggGraph)$
				\label{code:addtagend}	
			\ELSE
				\STATE $clusters$.newCluster($tag$, {$g$})
				\label{code:newcluster}
	    		\ENDIF		
		\ELSE
			\STATE $aggPlan$.add($tag$, $g$)
			\label{code:addplannode}
		\ENDIF
		\ENDIF
\ENDWHILE
\label{code:secondend}
\STATE aggregate($aggPlan$)
\label{code:aggregate}
\end{algorithmic}
\label{alg:aggregateplan}
\end{algorithm}

Given a set of $vertices$, we first generate tags for each vertex (Line~\ref{code:gentags}) then sort all tags and put into a queue based on their size and their values (Line~\ref{code:tagsort}). The benefit of this sorting operation is two-fold. First, after sorting, it is easier and fast to combine and pre-aggregate all the vertices with the same tag. Second, after sorting by size, we can guarantee that the larger tags can be clustered first.  
This is designed based on the fact that the longer tag it is, the larger possibility it has to provide a benefited sharing.

As the vertices with the same tags are definitely able to share their computation, for each popped tag in $queue$, vertices with same tags will be combined together into groups first(Line~\ref{code:combine}). This same tag combing is conducted until it reaches a different tag. Note that this coming is also a pre-aggregating procedure where the corresponding vertices information is pre-aggregated. 


After the first step of combing vertices with the same tags, we get a list of distinct tags each of which is associated with on group and the pre-aggregated value in the group. For instance, like in Figure~\ref{fig:sn-agg} (B), after the first combining step, B1 and B2 are combining into one group B with the tag $<1,2><3,4,5>$ and C1, C2 and C3 are into another group C with the tag $<1,2,3><4,5,>$.

In the second step, we discover more sharing opportunities among these distinct tags by clustering them into clusters according to their similarity. The general idea of this clustering procedure is as follows: Given a new tag, it compares all the existing clusters to find the best cluster which obtains the biggest saving value after adding the new tag into the cluster based on one saving function. The saving function will be provided in Equation~\ref{eq:saving}. If the biggest saving value is negative which means adding the new tag into any of the cluster does not increase the sharing opportunity, this new tag becomes a new cluster itself. This heuristic approach guarantees that the best cluster that increases the computation sharing is chosen in each clustering step. Since the tags are in sorted order, the clustering can stop while the new tag size becomes smaller than a threshold value, like 3. This is because most likely, when the tag size is smaller, the sharing opportunity is slightly small. There is no need to cluster them.

Now, we provide the saving equation used during the clustering. Assume for each cluster $C_i$, $CT_i$ is the common tag that is the intersection among all the tags in $C_i$. $SZ_i$ is the number of tags already in the cluster. Then the saving cost after adding a new tag $nt$ can be calculated as follows: 
\begin{equation}
\label{eq:saving}
saving(C_i,nt) = |CT_i \wedge nt| \times (SZ_i + 1) - |CT_i| \times SZ_i
\end{equation}
where 
$CT_i \wedge nt$ is the new common tag of the cluster after adding $nt$ to $C_i$, $ |CT_i \wedge nt| \times (SZ_i + 1)$  indicates the total saving of the new cluster after adding $nt$, $|CT_i| \times SZ_i$ indicates the aggregation saving of $C_i$ before adding the $nt$. Therefore the difference between these two costs are the benefit of adding a new tag to the cluster.

After the clustering, the aggregation can be conducted for each cluster.  Each tag $t$ in one cluster $C_i$ is actually split into two parts: one is the common tag $CT_i$ and another is the differential tag $DT_t$. Note that $DT_t$ of $t$ is the tag that is not covered by the common tag $CT_i$ of $C_i$. The $DT_t$ can be obtained by $T(t)-CT_i$. For instance, if $t$ is $<1><3,4,5>$ and $CT_i$ is $<1><3,4>$, $DT_t$ is $<1><5>$. For the common tag in each cluster, one further aggregation based on all the groups in one cluster can be conducted.  The aggregation results can be directly used to all the subgraphs indicating by the common tag. This saves the repeated aggregation among these group for each subgraph. 
For each member in the cluster, its pre-aggregate value from the first step needs to send to all the subgraphs representing in its differential tag.   

\begin{table*}[th]
\begin{center}
\caption{Aggregate performance over dense graph (ms)}
\label{tab:denseperformance}
\begin{tabular}{|l|c|c|c|c|c|c|c|c|c|c|}
\hline
\multirow{2}{*}{\backslashbox{C}{SV}} & \multicolumn{2}{c|}{5} & \multicolumn{2}{c|}{10} & \multicolumn{2}{c|}{20} & \multicolumn{2}{c|}{30} & \multicolumn{2}{c|}{40}
 \\ \cline{2-11}
& SN & AS  & SN & AS  & SN & AS & SN & AS  & SN & AS \\  \hline
10    & 1877  & 1383  & 2302  & 1233  & 3154  & 1009  & 4894  & 972   & 6175  & 990   \\
100   & 1888  & 1563  & 2433  & 1234  & 3432  & 969   & 5375  & 961   & 6862  & 1001   \\
1000  & 2110  & 1904  & 2823  & 1308  & 3828  & 1071  & 7281  & 1099   & 9264  & 1274 \\
10000 & 2656  & 2148  & 3084  & 1506  & 4454  & 1364  & 7696  & 1522   & 10367 & 1877 \\
100000 & 2927  & 2382  & 3192  & 1869  & 4370  & 2588  & 8403  & 2977  & 11452 & 3141 \\
200000 & 3028  & 2620  & 3215  & 2115  & 4608  & 2469  & 9174  & 3116  & 13100 & 4049 \\
400000 & \textit{3039}  & 3070  & \textit{3294}  & 3404  & 4664  & 3296  & 8587  & 6146  & 16380 & 5678 \\
600000 & \textit{2933}  & 3232  & \textit{3353}  & 3525  & 4782  & 3665  & 10018 & 5694  & 19798 & 7377 \\
800000 & \textit{3003}  & 3182  & \textit{3338}  & 3532  & \textit{4949}  & 5214  & 10227 & 6741  & 22852 & 8257 \\
1000000 & \textit{3113}  & 3361 & \textit{3497}  & 3691  & \textit{5372}  & 6082  & 10299 & 8000  & 24475 & 10151 \\
 \hline
\end{tabular}
\end{center}
\end{table*}

\begin{table*}[th]
\begin{center}
\caption{Aggregate performance over sparse graph(ms)}
\label{tab:sparseperformance}
\begin{tabular}{|l|c|c|c|c|c|c|c|c|c|c|}
\hline
\multirow{2}{*}{\backslashbox{C}{SV}} & \multicolumn{2}{c|}{5} & \multicolumn{2}{c|}{10} & \multicolumn{2}{c|}{20} & \multicolumn{2}{c|}{30} & \multicolumn{2}{c|}{40}
 \\ \cline{2-11}
& SN & AS  & SN & AS  & SN & AS & SN & AS  & SN & AS \\  \hline
10    & 502   & 418   & 625   & 375   & 664   & 229   & 963   & 227   & 1136  & 226  \\
100   & 513   & 428   & 665   & 377   & 708   & 231   & 1019  & 234   & 1357  & 244  \\
1000  & 548   & 464   & 682   & 401   & 776   & 276   & 1126  & 385   & 1544  & 345  \\
10000 & 573   & 516   & 713   & 446   & 843   & 360   & 1240  & 558   & 1604  & 540  \\
50000 & 587   & 584   & 537   & 389   & 923   & 581   & 1328  & 721   & 1677  & 1012  \\
100000 & \textit{593}   & 633   & 760   & 681   & 898   & 841   & 1284  & 1013 & 1775  & 1219  \\
150000 & \textit{602}   & 650   & 584   & 568   & 884   & 863   & 1340  & 1155 & 2221  & 1581  \\
200000 & \textit{645}   & 674   & \textit{763}   & 795   & 951   & 996   & 1305  & 1286 & 2508  & 1699  \\
\hline
\end{tabular}
\end{center}
\vspace*{-4mm}
\end{table*}

\section{EXPERIMENTAL EVALUATION}
\textbf{Environment.} We conduct all the experimental evaluations on a platform with an Intel Xeon E5607 4-core CPU (2.33GHz), 32GB of memory with running Linux 2.6.32 64-bit OS. 

\textbf{Implementation.} All algorithms are implemented using java. Transitive closure are used as reachability index to support the extraction of subgraphs.

\textbf{Datasets.} We perform our experimental studies on two kinds of datasets including one real Twitter dataset (provided by UIUC~\cite{UIUCDataset}) and a set of synthetic datasets. The Twitter dataset contains 284 million following relationships, 3 million user profiles and 50 million tweets. Each user profile has information about account age, location, etc, and Re-tweets contains information about origin, time, content, etc.

The synthetic datasets are generated using the GRAIL graph data generator. Each generated synthetic dataset is a directed attributed graph. Each vertex in the graph is associated with three attributes ($vid$, $v\_grp$, $v\_mr$) where $v\_grp$ and $v\_mr$ are the group and measure information with integer data type. Each edge is associated with four attributes ($src\_vid$, $tgt\_vid$, $e\_grp$, $e\_mr$), where $e\_grp$ and $e\_mr$ are edge group and measure information with integer data type as well. 

\subsection{Effectiveness}

We first show the effectiveness of VCExplorer as a powerful tool to explore the Twitter graph. 
Given the Twitter graph, we are interested in discovering who are the most active users and what are the distributions of  contact frequency among users in the influence subnetwork between them.  We use count of tweets between two users to compute their contact frequency. Bigger $frequency$ is,  stronger relationship they are. Further, we classify $frequency$ into three categories($Closeness$): High, Middle, and Low.
 
\begin{figure}
\includegraphics[width=3in]{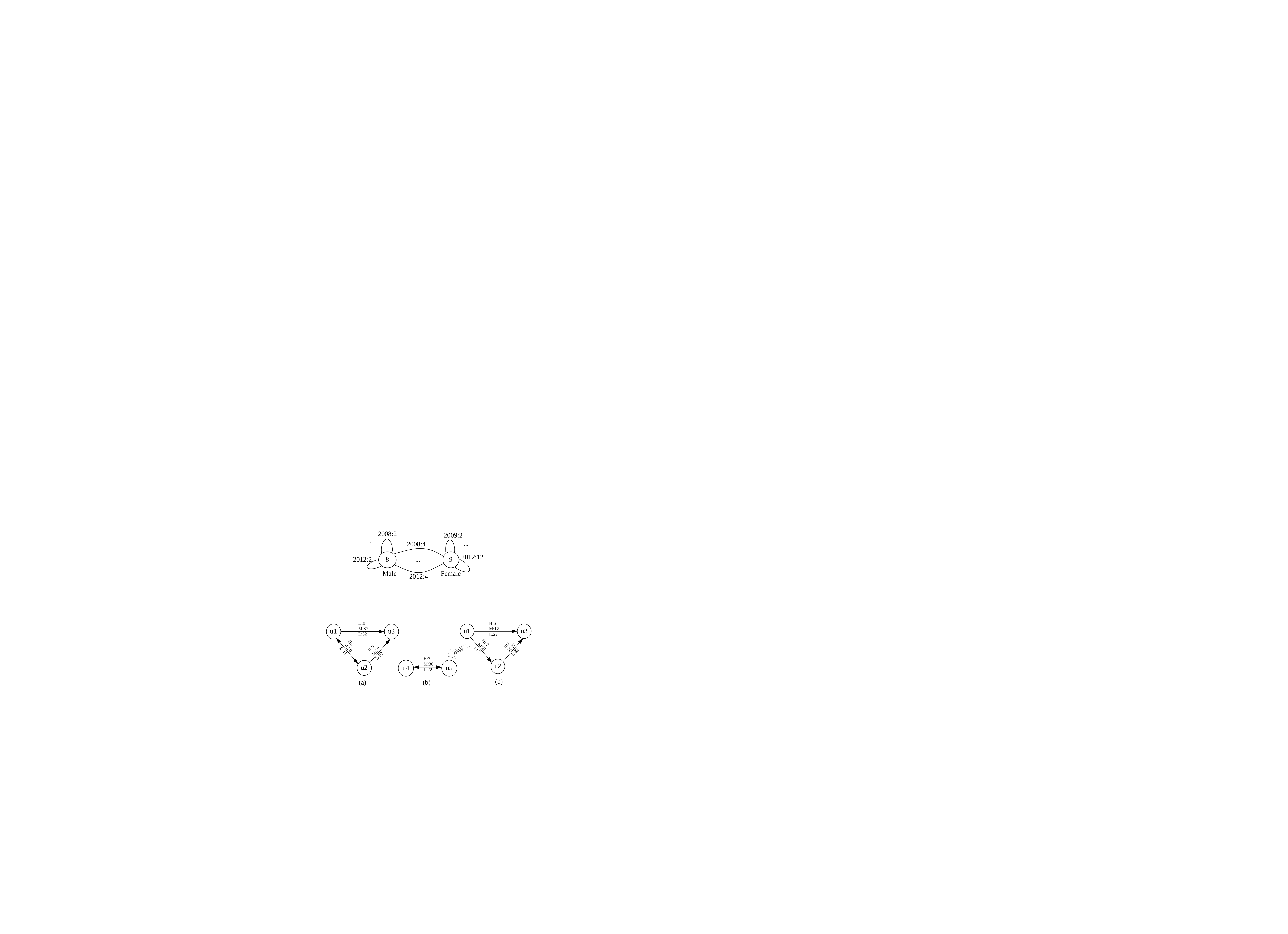}
\caption{HA-Graphs over Twitter Network.}
\label{fig:effectiveness}
\end{figure}

The  GE-query may be expressed as follows:
\begin{flushleft}\tt
\hskip 20pt \Sketch\ TopMaxDegreeVertices(twitter,3)\\
\hskip 20pt \From\ twitter \\
\hskip 20pt \Group\ betweeness() \\
\hskip 20pt \Summarize\ COUNT(.) e. Closeness()
\end{flushleft}

Resulting graph is shown in Figure~\ref{fig:effectiveness} (a). Distribution of different closeness categories of each subgraph are annotated on edge. From Figure~\ref{fig:effectiveness}(a) we may see that there is one circle between $u1$ and $u2$ which causes other edges ($u1$, $u3$) and ($u2$, $u3$) have the same distributions. So we change to another betweenness function to eliminate the circle affection: replace the $betweenness$ function with $betweenness(h)$ which check whether one vertex may reach another vertex within $h$ hops.  Figure~\ref{fig:effectiveness}(c) is the resulting HA-graph while $h=4$. One remarkable change is, high closeness relationships between $u1$ and $u2$ has been reduced from 7 to 2. Such remarkable change leads us to analyze the subnetwork between $u1$ and $u2$ deeply. We may issue a zoom query over subgraph between $u1$ and $u2$ with $k=2$ and $h=4$, zoom operation output a new HA-graph as shown in Figure~\ref{fig:effectiveness} (b). From the aggregate values on edge, it is easy to see that most strong relationships  between $u1$ and $u2$ are between $u4$ and $u5$, which indicates that middle users between $u4$ and $u5$ have stronger relationships.

\subsection{Performance Evaluation} 

In this section, we evaluate the performance of our proposed graph aggregation algorithm. Two algorithms are implemented and compared including the baseline algorithm \textbf{SN} - shared nothing algorithm as discussed in section~\ref{subsec:SN} and the \textbf{AS} algorithm - Aggregation Sharing algorithm as proposed in section~\ref{subsec:AS}. Note that all the following experiments are conducted three times and the average performance is reported.

The GE-query used is provided as follows: 

\begin{flushleft}\tt
\hskip 20pt \Sketch\ TopMaxDegreeVertices(k)\\
\hskip 20pt \From\ G \\
\hskip 20pt \Group\ betweeness() \\
\hskip 20pt \Summarize\ SumVMrByVGrpEGrp(), \\
\hskip 95pt SumEMrByVGrpEGrp() \\
\end{flushleft}

For simplicity, the GE-query used during the following experiments is to identify the top $k$ hub vertices with the maximum degree and summarize the relationship between two hub vertices by calculating the aggregate graph based on dimension v\_grp and e\_grp using SumVMrByVGrpEGrp() and SumEMrByVGrpEGrp() function which summarize v\_mr and e\_mr measures respectively by v\_grp and e\_grp.

Towards a comprehensive study, we study the impact of the number of hub vertices, graph dimension cardinality, graph degree and graph size accordingly. It is worthy of noting that the aggregation performance is affected by the cardinality of vertex group-by dimension and edge dimension together. These cardinalities will affect the final total different number of group-by values. Therefore, for simplicity, in the following experiments, we refer to the final total different number of group-by values of both vertices and edges as the cardinality.

\textbf{Impact of the number of hub vertices.} In this experiment, we first study the benefit of graph aggregation sharing over multiple sub-graphs when we vary the number of hub vertices (SV) from 5 to 40. We conduct the experiments over two different types of graphs: one with graph degree 8 representing a relative spare graph and another with degree 40 representing a relative dense graph. All the graphs used in these set of experiments consist of 30K vertices.

Table~\ref{tab:denseperformance} and Table~\ref{tab:sparseperformance} show detailed results for the graphs with degree 8 and 40 respectively. Note that each row indicates the execution time of different algorithms while selecting different number of hub vertices on the same graph with a specific cardinality showing the most left column. 

From the result, we have the following findings: First, SN and AS have different reactions when change SV. While SV increases, the execution time of shared nothing SN algorithm increases as well. The reason is that, as more hub vertices generate more influential subgraphs which leads more vertices and edges are involved into recomputation. Differently, AS does held this pattern. As shown in the result, while SV increases, the execution time does not increase as much. For some cases, it is even decreasing. For instance, in Table 1, the execution time of AS with SV=10 is always smaller than the one with SV=5 for smaller cardinality. This, however, is reasonable, as more hub vertices and smaller cardinality mean more sharing opportunities.

Second, as SV increases, AS outperforms SV more. As shown in Table 1 and 2, the execution time of SN increases dramatically while SV becomes larger. However, AS is more stable which leads AS outperforms SN more. 

\textbf{Impact of cardinality.} Table 1 and 2 also indicate how the performance changes when we vary the cardinality from 10 to 1,000,000. As expected, SN outperforms weakly AS only when the cardinality is large enough and SV is small. For instance, in Table 1, when SV=5, SN becomes faster than AS when the cardinality reaches 400,000(Italic numbers). This is because a larger cardinality reduces the opportunity of sharing operation. 

\begin{figure*}[th]
\centering
\begin{subfigure}[b]{0.5\textwidth}
\includegraphics[width=3in]{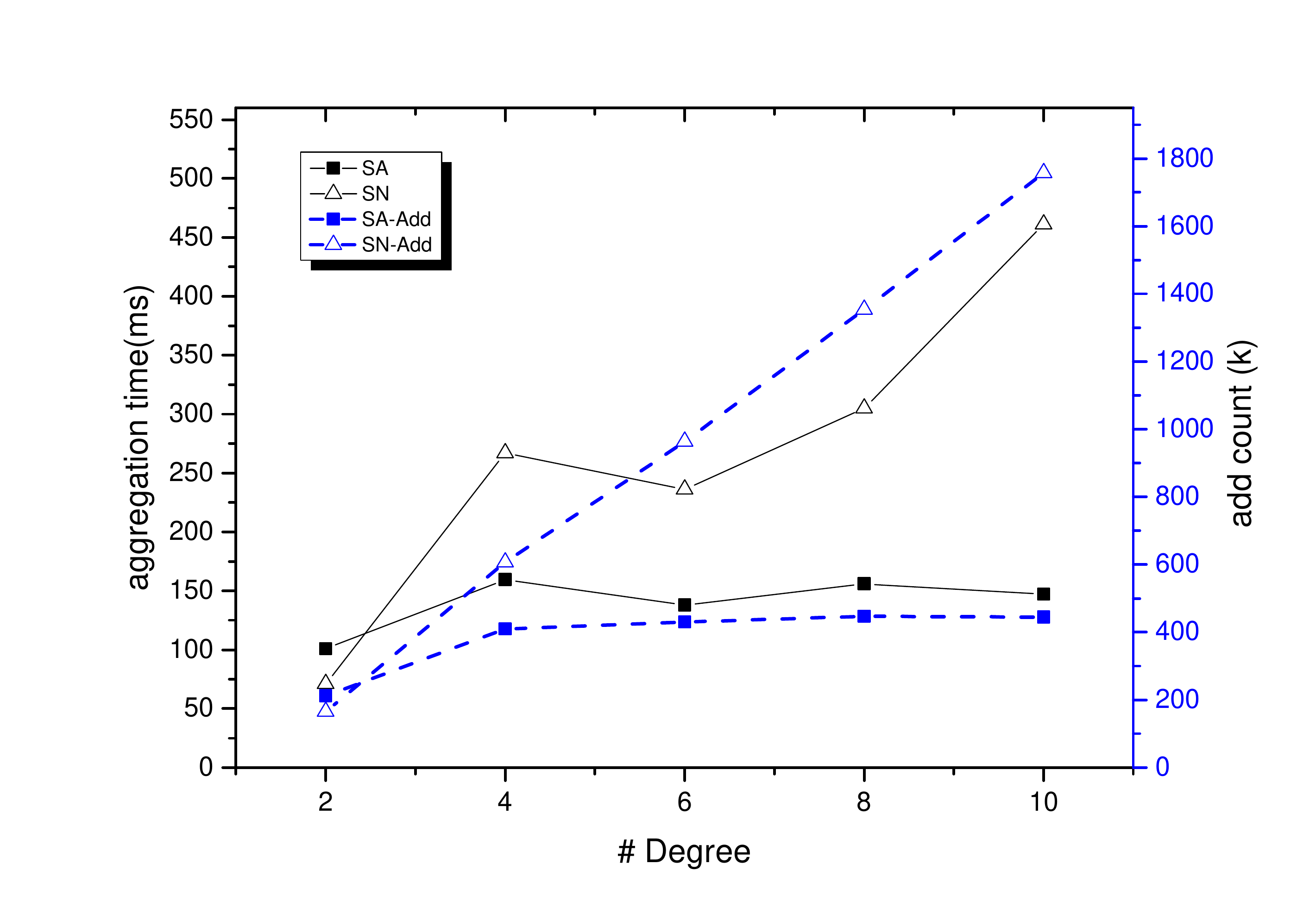}
\caption{Sparse graph.}
\label{fig:degree-sparse}
\end{subfigure}%
\begin{subfigure}[b]{0.5\textwidth}
\includegraphics[width=3in]{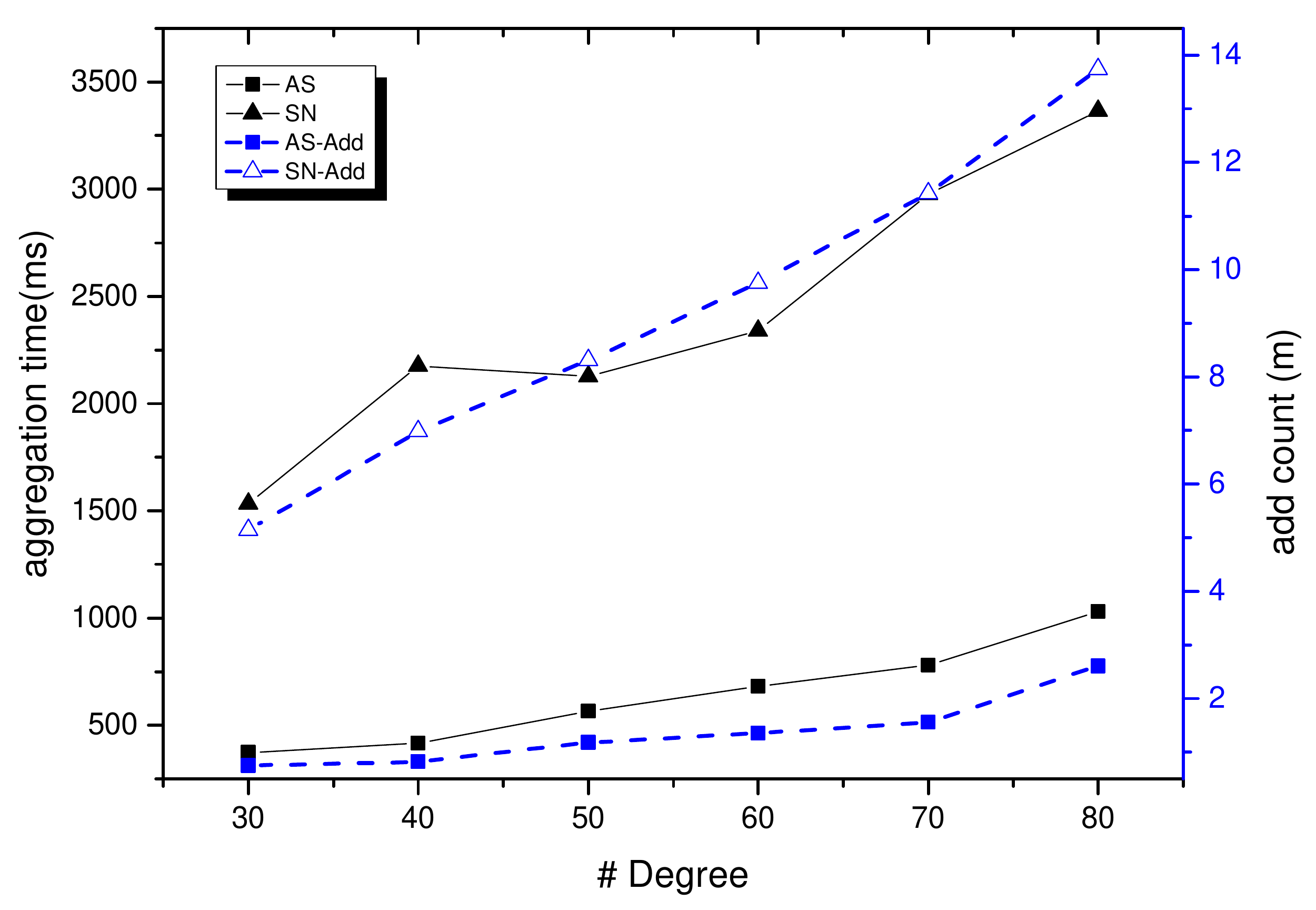}
\caption{Dengse graph.}
\label{fig:degree-dense}
\end{subfigure}
\vspace*{0mm}
\caption{Scalability vs graph degree.}
\vspace*{0mm}
\label{fig:scalabilitydegree}
\end{figure*}

\begin{figure*}[th]
\centering
\begin{subfigure}[b]{0.5\textwidth}
\includegraphics[width=3in]{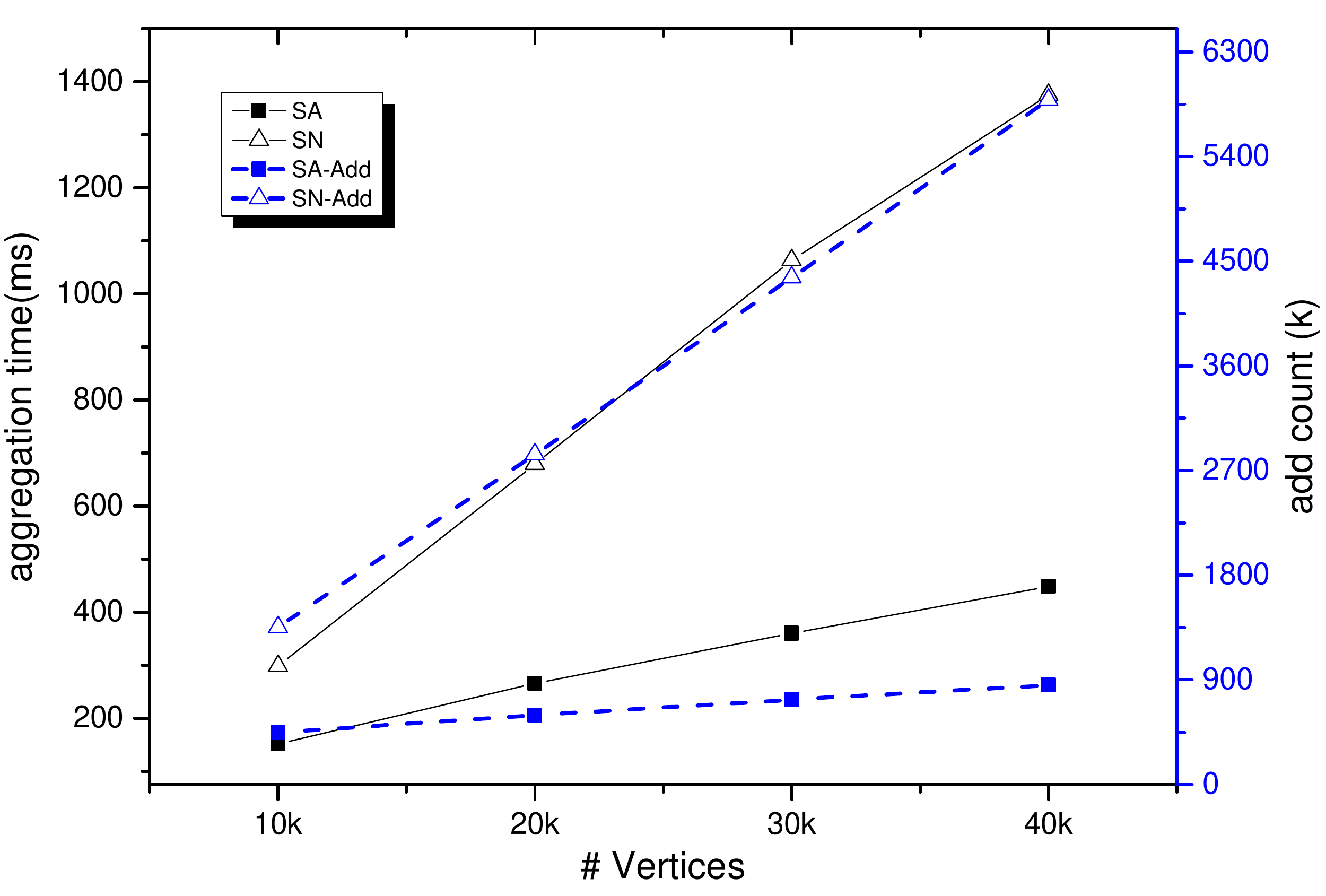}
\caption{Sparse graph.}
\label{fig:vertex-sparse}
\end{subfigure}%
\begin{subfigure}[b]{0.5\textwidth}
\includegraphics[width=3in]{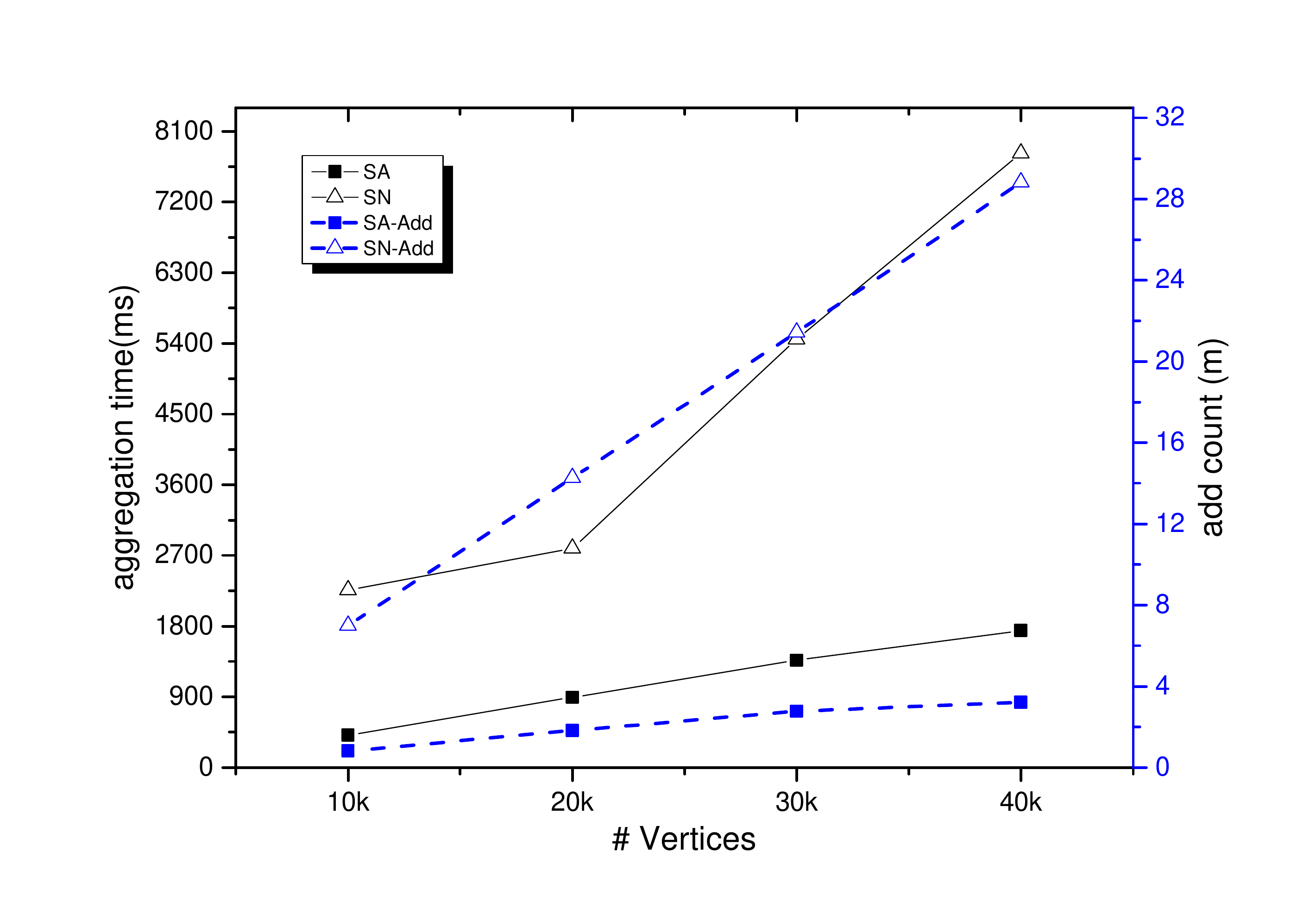}
\caption{Dense graph.}
\label{fig:vertex-dense}
\end{subfigure}
\vspace*{0mm}
\caption{Scalability vs graph size.}
\vspace*{0mm}
\label{fig:scalabilityvertex}
\end{figure*}

\begin{figure*}[!htb]
\centering
\begin{subfigure}[b]{0.3\textwidth}
\includegraphics[width=2.2in]{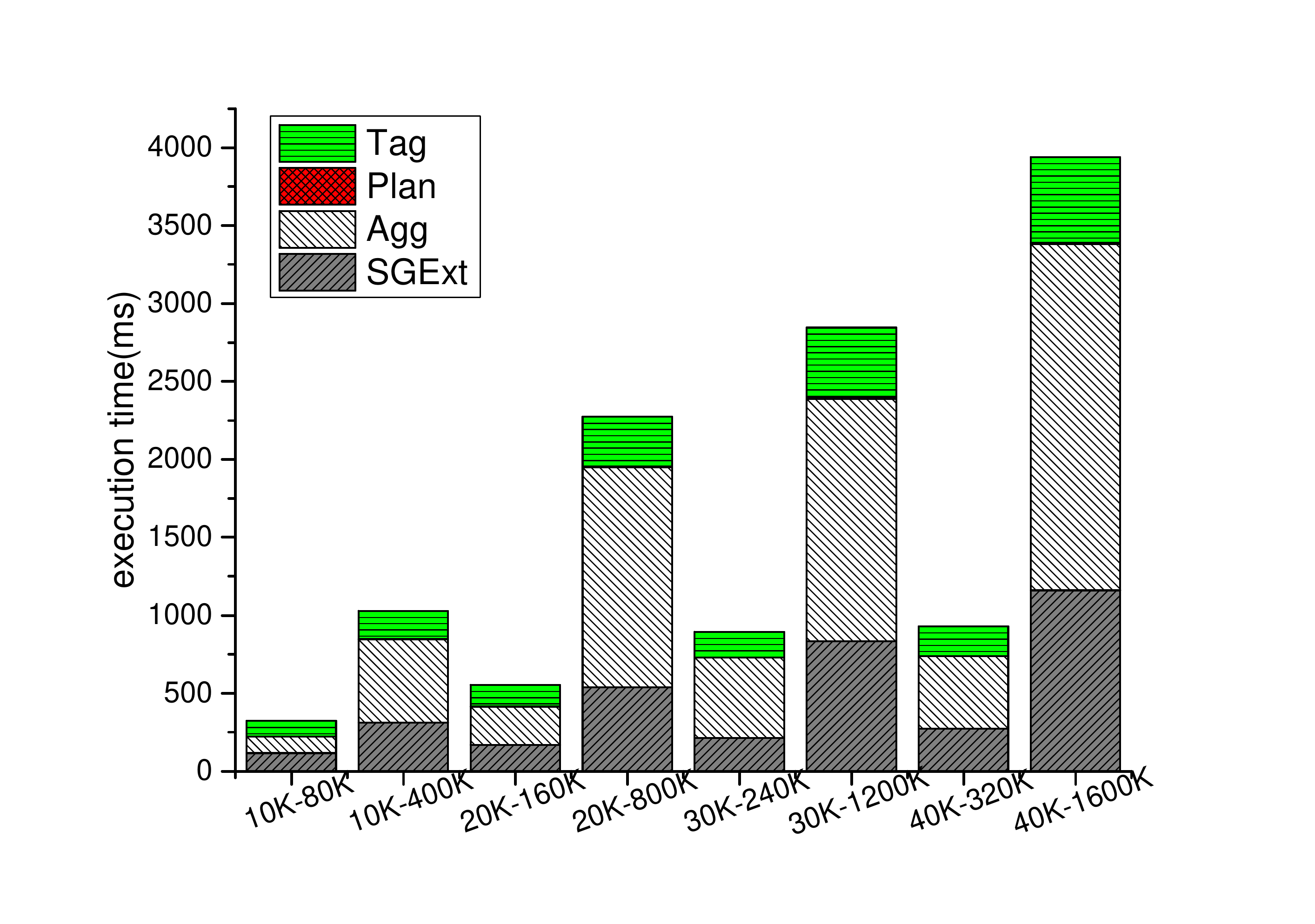}
\caption{SV=5.}
\label{fig:planning}
\end{subfigure}%
\begin{subfigure}[b]{0.3\textwidth}
\includegraphics[width=2.2in]{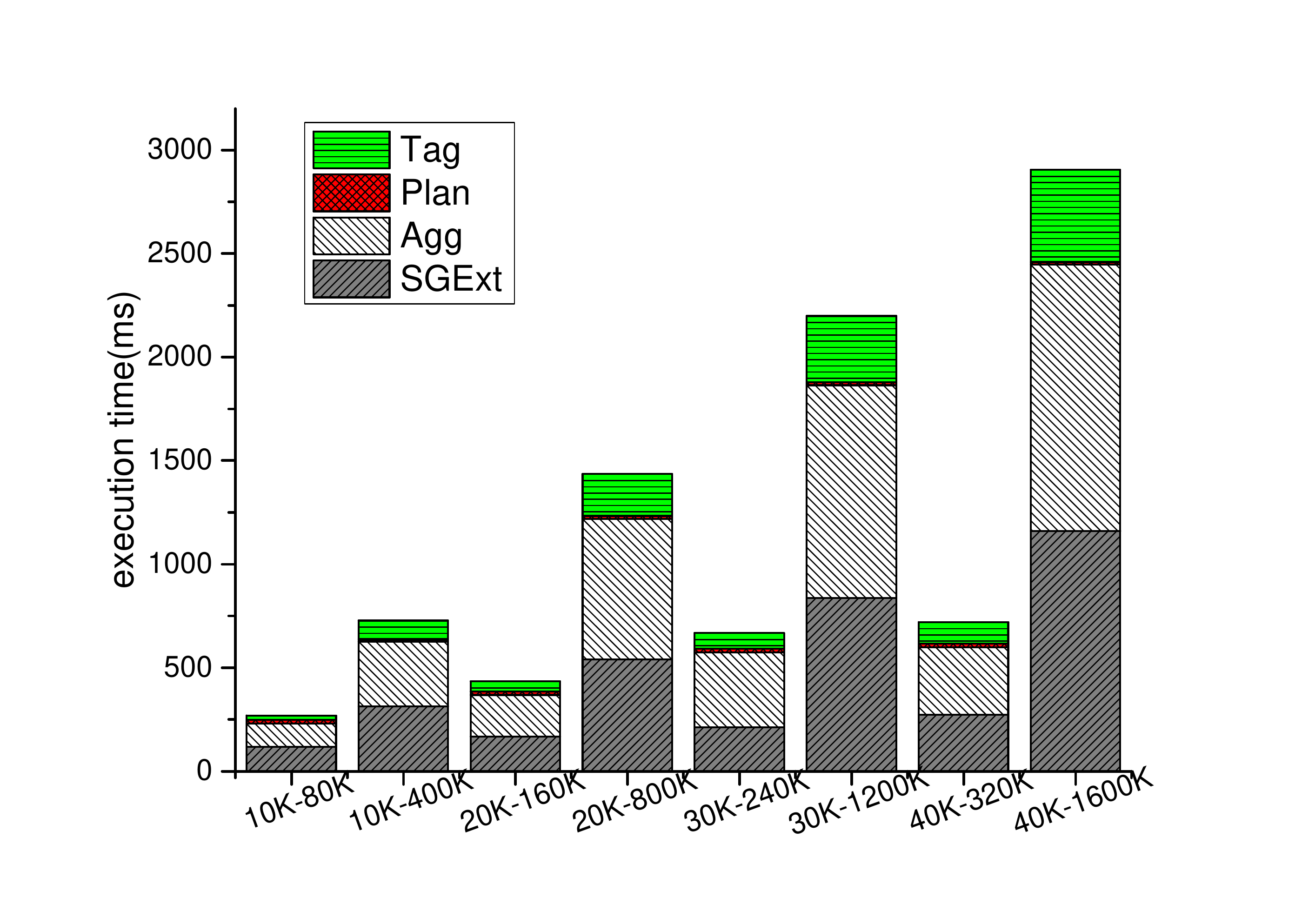}
\caption{SV=20.}
\label{fig:planning}
\end{subfigure}%
\begin{subfigure}[b]{0.3\textwidth}
\includegraphics[width=2.2in]{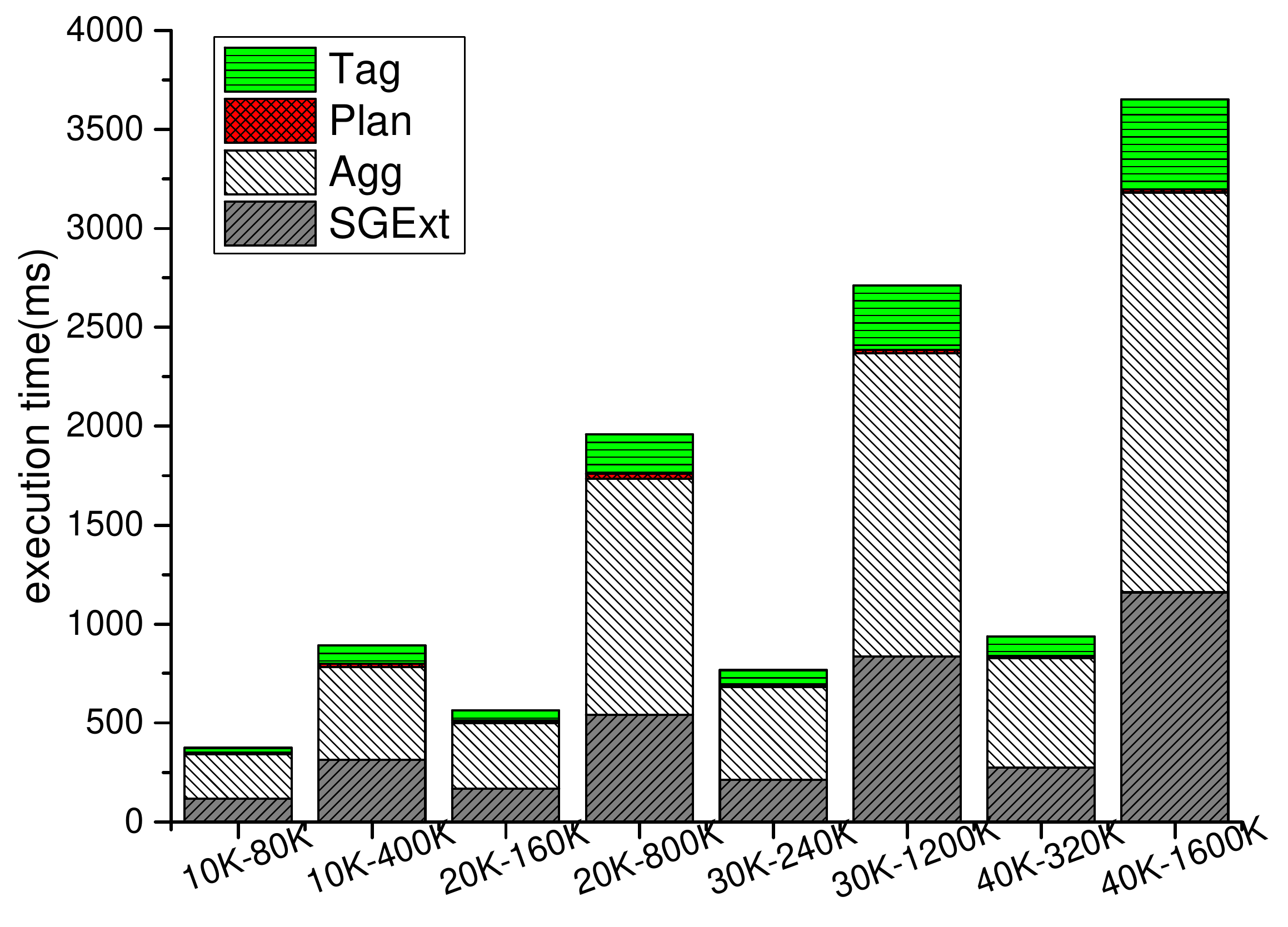}
\caption{SV=40.}
\label{fig:planning}
\end{subfigure}
\vspace*{4mm}
\caption{Time distribution for AS.}
\vspace*{0mm}
\label{fig:timedistribution}
\end{figure*}

\textbf{Impact of graph degree.} In this set of experiments, we study the performance comparison among different algorithms while we change the graph degree from 2 to 80. These experiments are conducted with SV=20 and C=10K based on the graphs with 10,000 vertices.  

Figure~\ref{fig:scalabilitydegree} (a) and (b) show the execution time(solid line) for relative sparse graph and dense graph respectively. From the result, we can see that as the degree increases, the query execution time of all the algorithms increases as well. It also indicates that AS is more stable than SN. 

To better understand, how many add operations are saved by the sharing algorithm. We collect the total number of add operations and show them as dash line in Figure~\ref{fig:scalabilitydegree}. From Figure~\ref{fig:scalabilitydegree}, we can see that the reason the AS can outperform SN dramatically is because it saves many add operations by sharing. In average, AS saved 74\%  and 60\% add operations in dense graph and sparse graph respectively compared to SN. 

\textbf{Impact of the number of vertices.} In this set of experiments, we study how the performance changes while we fix the graph degree but vary the number of vertices from 10K to 40K. 

Figure~\ref{fig:scalabilityvertex} (a) and (b) provide the execution time (solid line) based on the graphs with degree 8 and 40 each of which represents relative sparse or dense graphs. The results indicate that the execution time of SN algorithm increases faster than the execution time of AS algorithm when the number of vertices is increased. We further calculate the number of add operations incur in each experiment as shown as dash line in Figure~\ref{fig:scalabilityvertex}. It is easy to see that the number of add operations in SN algorithm becomes much larger than the ones in AS algorithm. These experiments also show that both SN and AS scale linearly when vertex number increases. 

\textbf{Time Distribution Analysis.}
To better understand the proposed AS algorithm, we run a set of experiments and count the running time of each part, including tagging generation (referred as Tag), subgraph extraction (referred as SGExt), planning time (referred as Plan) and aggregation time (referred as Agg). The experiments are conducted over a set of graphs with C=10,000 by running the query with different number of hub vertices (5, 20 and 40).

Figure~\ref{fig:timedistribution} (a), (b) and (c) present the overview of query execution time distribution with SV=5, 20 and 40 respectively. Note that the x axis indicates the graph size used. For instance, 10K-80K means the graph consists of 10K vertices and 80K edges. 40K-1600K means the graph consists of 40K vertices and 1600K edges and so on. The results indicate that the planning is very fast compared with other operations. The tagging time and subgraph extraction time occupy about 13\% and 38\% of total query time in average respectively. In whatever cases, the aggregation time takes the big portion of the total execution time.

\section{Related Work}
A great challenge in graph analytic is to deal with the presence of large attributed graphs. Related work on graph analytic can be summarized as follows:

\textbf{Graph Layout Drawing} aims to display whole graph in a user friendly way. Classic graph drawing algorithms are surveyed in \cite{herman2000graph}. Those algorithms can structurally display small graph on the screen. In order to enable user discerning on interesting vertices and edges, some discriminating methods are proposed in the literature. \emph{Position discriminating} methods \cite{Brandes2003, Merrick2006} place vertices with high centrality  \cite{sabidussi1966centrality,freeman1977set,noah1991theoretical} near the center of graph. Some other works \cite{bastian2009gephi} use \emph{Size discriminating} methods by displaying vertices with high importance value in larger circles or using prominent colors \cite{cleveland1984graphical}.
All these algorithms suffer from the volume of graphs. When graph size is up to tens of thousands vertices and edges, the screen will be filled up with dots, and the link information among vertices is barely seen. In contrast, our SVExpolorer displays sketch graph which contains less vertices and consolidated information between vertices. By so doing, user will not get overwhelming points in the display. 

\textbf{Graph Simplification} aims to reduce graph size prior to above layout algorithms. Several approaches are developed for this purpose. \cite{abello2006ask, archambault2007grouse} group strongly connected vertices and edges into metanodes. \cite{ellis2007taxonomy} merges edges in the same simple path or routes, \cite{dickerson2004confluent,holten2006hierarchical} condense non-planar graph into planar graphs, and \cite{gansner2007improved,dwyer2007integrating,ersoy2011skeleton} form edge bundles by some metrics. \cite{higbee1998mathematical,schaffer1996navigating} uses clustering based approach to form hierarchical view of the graph, which supports navigation. ~\cite{PientaKLVTAPC17} reduces graph size by displaying only nodes and neighborhoods that are most subjectively interesting to users. However, all these methods cannot handle attributed graphs as in our case. First, since vertices and edges to be retained in the simplification algorithm is selected automatically, users are not feasible to choose particular points and view the relationships among them. Second, most of these methods only concern the structure of graphs, the attributes of vertices and edges are not preserved. On the contrary, VCExplorer enables users arbitrarily picking of the interesting vertices, and further provides consolidated information  among these vertices.

\textbf{Graph Summarization} aims to provide a succinct high-level graph by consolidating vertex's attributes and edge information. Vertices and edges belonging to the same metric are viewed as metanodes and edges. Aggregated information from detail vertices and edges are attached to the metanodes and edges. \cite{Tian2008} develops k-SNAP method for cluster graph into k groups. \cite{Zhao2011, wangzk2014} has proposed graph aggregation methods which group the graph based on vertex and edge attributes. These methods offer good overview of graph attributes in a succinct way, but they do not position important vertices and their relationships. Although~\cite{Miao2017} summarizes graph according to the importance and relatedness of vertices, it focus mainly detailed vertices. Differently, our VCExplorer displays important vertices as hub vertices and reveals the relationships between them using  consolidation techniques.



\end{document}